
\input amstex
\magnification=1200
\TagsOnRight
\def\wh{\widehat}

\def\D{\Cal{D}}

\def\noi{\noindent}

\overfullrule=0pt

\def\mapright#1{\smash{\mathop{\longrightarrow}\limits^{#1}}}
\def\mapab#1{\Big\downarrow}

\mathsurround=1pt
\tolerance=10000
\pretolerance=10000
\nopagenumbers
\headline={\ifnum\pageno=1\hfil\else\hss\tenrm -- \folio\ --\hss\fi}
\hsize=16 true cm
\vsize=23 true cm

\line{Preprint {\bf SB/F/94-226}}
\hrule
\vglue 1.5cm
\vskip 2.cm

\centerline{\bf BRST QUANTIZATION OF NON-ABELIAN }
\centerline{\bf BF TOPOLOGICAL THEORIES  }

\vskip 1.5cm
\centerline{ M. I. Caicedo, R. Gianvittorio, A. Restuccia and J. Stephany}
\vskip .5cm
\centerline{\it Universidad Sim\'{o}n Bol\'{\i}var, Departamento de
F\'{\i}sica}
\centerline{\it Apartado postal 89000, Caracas 1080-A, Venezuela.}
\centerline{\it e-mail: mcaicedo{\@}usb.ve, ritagian{\@}usb.ve,
arestu{\@}usb.ve, stephany{\@}usb.ve}
\vskip 1cm
{\bf Abstract}
\vskip .5cm
{\narrower{\flushpar  The off-shell nilpotent BRST charge and the BRST
invariant effective action for non-abelian BF topological theories over
D-dimensional manifolds  are explicitly constructed. These theories have the
feature of being  reducible  with exactly D-3
stages of reducibility. The adequate extended phase space including the
different levels of ghosts for ghosts is explicitly obtained. Using
the structure of the resulting BRST charge  we show that for topological BF
theories the semi-classical approximation completely describes the quantum
theory. The independence of the partition function on the metric also follows
from our explicit construction in a straightforward way. \par}}

\vskip 3cm
\hrule
\bigskip
\centerline {\bf UNIVERSIDAD SIMON BOLIVAR}

\newpage

BF Topological actions which were introduced in [1,2] as generalizations of
three
dimensional Chern Simons theories,  can also be regarded as a zero
coupling limit of Yang-Mills theories. Some results concerning their
quantization
in the abelian case are fairly well understood, indeed, it has been shown that
the partition function for the abelian case may be written in terms of  the Ray
Singer torsion [3] while other observables such as the Wilson surfaces
determine
linking and intersection numbers of manifolds in any dimensions.

For the non-abelian case, the metric independence of the quantum  BF theory was
first
proposed in [3] and proved later on in references [4-6] in a direct way. A
solution of the master equation of the Batalin-Vilkoviski (BV) approach for the
BF
action was also presented in [7]. The authors of reference [8] were able to
build an off-shell nilpotent Becchi-Rouet-Stora-Tyutin (BRST) operator using an
approach similar
to the BV procedure but with a Slavnov identity playing the role of the
BV master equation. The construction presented in [8] was performed in a
covariant
gauge and enabled the authors to show that perturbatively the 4 and 5
dimensional
theories are anomaly free and finite.

The construction of an off-shell
nilpotent BRST charge for a non abelian BF theory
based on a canonical (symplectic) formalism was first carried out for the four
dimensional model in reference [9].
This operator was built following a modified BFV method
introduced earlier in refs. [10] and [11],
which simplifies some aspects of the usual approach [12]. Several issues
concerning the general structure of of the off shell BRST invariant action for
topological theories and in particular a general proof of the metric
independence of the partition function  were also commented in [9].

In the present paper we explicitly construct
the off shell BRST invariant effective action for the non abelian BF theories
in
any dimension and for any admissible gauge fixing condition.
Using this effective action, we prove that in
any BF theory the partition function is independent of the gauge coupling
constant ($e^2$), a feature which also
appears in Witten's Topological Field Theory [13-16], and which has been
recently used to
obtain new topological invariants for four dimensional manifolds.
It is interesting to notice that Seiberg and Witten\' s [17] new approach is
based on
the structure of N=2 supersymmetric theories in the strong coupling regime
and on the independence of the topological theory in the gauge
coupling constant properties which are also present in BF theories.

The metric independence
of the partition function for these systems arises from general arguments.
Although metric dependent terms may appear in the effective action through
the gauge fixing conditions, the independence of the partition function
on them and consequently on the metric is guaranteed by virtue of the BFV
theorem [10-12].

The main difficulties in working with BF actions are related to the fact that
they are reducible theories, this property means that many levels of
ghosts for ghosts are needed in order to build the proper extended phase space.
The BF actions defined on D-dimensional manifolds
have exactly D-3 levels of reducibility [3] which make them particularly
interesting since such property turns them into explicit examples of reducible
theories with a higher ($\geq 2$) but finite number of stages of reducibility.

The action of a BF theory on a principal bundle with a D dimensional base
manifold is given in terms of the curvature 2 form ($ F \equiv
[ \nabla , \nabla] $) and a Lie algebra valued D-2 form ($ B
$) as follows:

$$
S={1\over{4}}\int Tr(B\wedge F) ={1\over{4}}\int
d^Dx \
\epsilon^{\mu_1\cdots\mu_D}(B_{\mu_1\cdots\mu_{D-2}}^a)
(F_{\mu_{D-1}\mu_D}^a).\tag1
$$

For the sake of clarity we will begin by discussing the
five dimensional case since it presents some features that are absent in four
dimensions and which constitute the clue for the discussion of theories
formulated in higher dimensions ($D\ge{}5$)
According to (1), the 5 - dimensional BF theory is given by

$$
S={1\over{4}}\int d^5x \
\epsilon^{\mu\nu\alpha\rho\sigma}(B_{\mu\nu\alpha}^a)(F_{\rho\sigma}^a).\tag2
$$

After integration by parts and defining $\epsilon^{ijk}\equiv
\epsilon^{0ijk}$, the original action (2) may be rewritten as

$$ S=\int
d^5x \
\dot{A}_i^a[{1\over{2}}\epsilon^{ijkl}(B_{jkl}^a)]+A_0^a\nabla_i
[{1\over{2}}\epsilon^{ijkl}(B_{jkl}^a)]+B_{0ij}^a({1\over{2}}\epsilon^{ijkl}F_{kl}^a).
\tag 3  $$

This expression can be clearly recognized as a canonical action for the
simplectic pairs $(A_i^a,\pi^{ia})$
where
$$
\pi^{ia}={1\over{2}}\epsilon^{ijkl}(B_{jkl}^a) .
$$

The action (3) has a vanishing Hamiltonian, and is subjected to the following
set of primary constraints:

$$
\align
&\phi^a \equiv
(\nabla_i\pi^i)^a=\partial_i\pi^{ia}+(A_i\times
\pi^i)^a=\partial_i\pi^{ia}+f^{abc}A_i^b\pi^{ic}=0 ,\tag 4a\\
&\Phi^{ija} \equiv \epsilon^{ijkl}(F_{kl}^a)=0 ,\tag 4b
\endalign
$$
whose associated Lagrange multipliers are $A_0^a$ and $B_{0ij}^a$.

The Poisson bracket algebra of the above constraints is given by:

$$
\align
&\{\phi^a(x),\phi^b(x')\} =f^{abc}\phi^c(x)\delta^3(x-x'),\tag5a \\
&\{\Phi^{ija}(x),\Phi^{klb}(x')\} =0, \tag5b \\
&\{\phi^a(x),\Phi^{ijb}(x')\} =f^{abc}\Phi^{ijc}(x)\delta^3(x-x'),\tag5c
\endalign
$$
showing that they are first class. It is important to notice that the set
of scalar constraints $\phi^a(x)$ constitutes a closed irreducible sub-algebra
of the whole set of constraints while the remaining constraints, namely the
tensor ones $\Phi^{ijc}(x)$ are linearly dependent i.e. reducible, since they
satisfy the identity:
$$
(\delta^m_{[i} \nabla_{j]}\Phi^{ij})^a=0 .\tag 6
$$

Moreover, this particular algebra is two times reducible since the
matrix $a^{(1)}\equiv a^{(1) ma}_{ij}= \delta^m_{[i}\nabla^a_{j]}$ is
itself reducible through the action of the following operator:
$$
a^{(2)}\equiv a^{(2) a}_{i}= \nabla^a_{i} . \tag7
$$
This second stage of reducibility holds only on-shell. Indeed, one can easily
check that:
$$
a^{(2) a}_{m} a^{(1) mb}_{ij} = \nabla^a_{m} \delta^m_{[i} \nabla^b_{j]}=
\nabla^a_{[i} \nabla^b_{j]} \sim f^{abc}\Phi^{ijc} \approx 0 . \tag8
$$
This last feature will be relevant when calculating the BRST charge
($\Omega$), since in order to have a manifest BRST invariant quantum theory
$\Omega$ must be off shell nilpotent.

In the modified BFV approach we are using, the extended phase space of a
reducible
sistem is divided in   a minimal sector of canonical pairs with
canonical BRST transformation laws  and a non-minimal sector. The minimal
sector is composed by the original fields ($q,p$) and a chain of
as many pairs of conjugate ghosts and ghosts for ghosts as levels of
reducibility. The
non minimal sector is composed by  extra ghosts, anti-ghosts and Lagrange
multipliers   whose transformation laws are deviced in a way that
ensure the off-shell closure of the BRST transformation [10-12].

  For this system, due to the fact that it has two stages of reducibility, the
minimal sector of the phase space is composed by the following canonical
pairs:

The original fields
$$
(A_i^a,\pi^{ia}) , \tag 9a
$$
the ghosts associated with the irreducible constraints and their momenta
$$
(C_1^a,\mu^{1a}) , \tag 9b
$$
and finally, the tower of ghosts and ghosts for ghosts associated
to the reducible constraints and their corresponding conjugate momenta
$$
(C_{1ij}^{(0)a},\mu^{1ija}_{(0)});(C_{11i}^{(1)a}
,\mu^{11ia}_{(1)}),(C_{111}^{(2)a},\mu^{111a}_{(2)}) .
\tag 9c
$$
Here and in the rest of this paper an embraced super(sub)script indicates
the level of reducibility to which a field is associated. Fields associated
to irreducible constraints carry no such label. The non minimal sector is
discused below.

 The BRST charge is constructed using only the variables of the minimal
sector.  A first candidate for the BRST charge is given by:
$$ \align
\Omega^{(naive)}=<&C_1 \phi - {1\over{2}}C_1(C_1 \times \mu^1) \\
& + C_{1 ij}^{(0)} \Phi^{1 ij} + C_{11 m}^{(1)} \delta^m_{[i}\nabla_{j]}
\mu^{1ij}_{(0)} + C_{111}^{(2)} \nabla_{j} \mu^{11 j}_{(1)} \\
&-C_1(C_{1 ij}^{(0)} \times \mu^{1 ij}_{(0)}) - C_1(C_{11m}^{(1)} \times
\mu^{11 m}_{(1)})
- C_1(C_{111}^{(2)} \times \mu^{111}_{(2)})> .\tag 10
\endalign
$$
where $C_1(C_1 \times \mu^1) \equiv
f^{abc}C_1^a C_1^b \mu^{1c}$ ,etc. and $< \cdots >$ stands for integration on
the space like continuous indexes.
This charge has the same basic structure discussed in
reference [9] on which we comment briefly. The first line in
$\Omega^{(naive)}$ looks like in standard Yang-Mills. This is not surprising
since it corresponds to the irreducible sub-algebra (5a). The second line
comprises terms coming from the reducible sector of the constraints (4b) and
finally, the last line, which is again Yang-Mills like, is due to those
Poisson brackets that mix both sectors of the algebra (5c). This structure
looks sufficiently rich for obtaining a nilpotent BRST charge but the direct
computation results in

$$
\{ \Omega^{(naive)} ,\Omega^{(naive)} \} \sim \{ C_{11 m}^{(1)}
\delta^m_{[i}\nabla_{j]} \mu^{1ij}_{(0)}, C_{111}^{(2)} \nabla_{j} \mu^{11
j}_{(1)} \} \approx
\Phi_{ij} C_{111}^{(2)} \mu^{1ij}_{(0)} , \tag11
$$
obviously meaning that the naive BRST charge is not off-shell nilpotent  (a
behavior that is not found in D=4) hence, we need  terms of higher order
in the ghosts to annihilate the unwanted term (11). At this stage, the only
contribution  that we can add which is compatible with both the ghost  and
geometrical structure of the minimal sector of the extended phase space is
given by the following trilinear combination of $C's$ and $\mu's$:
$C_{111}^{(2)}(\mu^{1ij}_{(0)} \times \mu^{1kl}_{(0)}) \epsilon_{ijkl}$
according to which the $\Omega$ generator should be written as

$$
 \Omega =  \Omega^{(naive)} + \alpha C_{111}^{(2)}(\mu^{1ij}_{(0)} \times
\mu^{1kl}_{(0)}) \epsilon_{ijkl} , \tag12
$$
where $\alpha$ must be calculated to ensure the off-shell closure of the BRST
algebra. One can in fact do this and obtain

$$
\alpha = -{1\over{8}} . \tag13
$$

{}From formulas (8) and (10) and according to comments given above, one
realizes
that the new term in the $ \Omega $ generator comes from the fact that the
second reducibility condition holds only on-shell
($ a^{(2)} a^{(1)}\sim \Phi \approx 0 $). We will see that in the higher
dimensional case this feature
reappears producing a series of polynomial terms in the BRST charge which
are generalizations of the trilinear term appearing in (12). There are two
important points  to remark about the BRST charge that we have just
calculated: the first is the fact that (12) is nilpotent off-shell, the
second is that $\Omega$ is linear in all the  ghosts associated to the
reducible sector,
(i.e. in: $C_{1ij}^{(0)} , C_{11i}^{(1)}$ and $C_{111}^{(2)}$). This latest
fact (which generalizes
for $D\ge 6$) is the key ingredient in the proof of the independence of the
partition function on the gauge coupling constant.

Let us now consider the non-minimal sector of the phase space, the auxiliary
fields belonging to this set are necessary in order to build the effective
action for
the theory and have been already described in reference [10]. These fields
include
extra ghosts, antighosts and Lagrange multipliers all of
which are Lie algebra valued.

In first place in table 1 we will introduce the complete set of
auxiliary fields associated with the irreducible constraints (4a)
\vskip 1cm
$$
\matrix
(C_{1},\mu^{1})&C_{2}&C_{3} \\
\lambda_{1}&\theta_{1} \\
\lambda_{11}&\lambda_{12}&\lambda_{13} \\
\endmatrix
$$
\centerline {{\bf Table 1:} Full set of irreducible auxiliary fields for the
$D=5$ BF theory.}
\vskip 1cm

In second place we introduce the set of auxiliary fields associated to the
reducible constraints, which as in the former case consists of two sectors: one
containig $C$ and $\mu$ fields and one containing extra lagrange mupltipliers:
the $\lambda$ and $\theta$ fields. The structure of the complete (both minimal
and nonminimal) set of $C$ fields can be graphically organized in a tree like
diagram one of whose branches we show in figure 1.
\vskip 1cm
$$
\matrix
C^{(0)}_{N} \\
\mapab{} \\
C^{(1)}_{N1}\mapright{}
&\mapright{}&\mapright{}&C^{(1)}_{N2}\mapright{}
&C^{(1)}_{N3} \\
\mapab{}&&&\mapab{}&\mapab{} \\
C^{(2)}_{N11}\mapright{}
&C^{(2)}_{N12}\mapright{}
&C^{(2)}_{N13}{}&\vdots{}&\vdots{}\\
\endmatrix
$$
\vskip .5cm
\centerline {{\bf Figure 1}: Reducible $C$ fields for the $D=5$ Topological
theory. $N=1,2,3$}
\vskip 1cm

In the diagram above one has to remember that the objects in the minimal sector
i.e.
$C^{(1)}_{1}$, $C^{(1)}_{11}$ and $C^{(2)}_{111}$ must be accompanied by their
conjugate
momenta. In this notation and according to [10], the bracketed superscripts
constitute a bookkeeping device that tells to which stage of reducibility does
a particular object belong.

The $\lambda$ and $\theta$ fields are many more than the former $C$ fields,
nevertheless they may also be organized as a family of
layers corresponding to each level of reducibility. For these fields one needs
an extra
superindex which labels the origin of the object. The complete family is
displayed in
figure 2 below.
\vskip 1cm
$$
\matrix
\lambda_{1}^{0(0)},\theta_{1}^{0(0)}&&\lambda_{1M}^{0(1)},
\theta_{1M}^{0(1)}&&\lambda_{1MN}^{0(2)},\theta_{1MN}^{0(2)} \\
&&\lambda_{11}^{1(1)},\theta_{11}^{1(1)}&&\lambda_{11M}^{1(2)},\theta_{11M}^{1(2)} \\
&&&&\lambda_{111}^{2(2)},\theta_{111}^{2(2)} \\
\endmatrix
$$
\vskip .5cm
\centerline {{\bf Figure 2}: Reducible $\lambda$ and $\theta$ fields for the
$D=5$ BF theory
, $M,N=1,2,3$}
\vskip 1cm

In the matrix like structure shown in figure 2 each row defines two
independent tree diagrams (one for the $\lambda$ and
one for the $\theta$ fields) like those ones defined in figure 1. In each
field the label appearing before the bracketed superscript tracks the
level in which each tree begins while as stated before, the bracketed
superscript
defines the stage of reducibility to which the object does actually belong.
For example, $\lambda^{1(2)}_{11M}$ is a second stage auxiliary field belonging
to
a tree begining at the first stage of reducibility. Notice that the arrengement
is upper triangular, meaning that there
are no ghosts either for $\lambda^{2(2)}_{111}$ or $\theta^{2(2)}_{111}$.

All these auxiliary fields are needed to ensure both the off-shell nilpotency
of the BRST transformations and the construction of a BRST invariant effective
action. The  capital subscripts (identifiers) that appear in each field
are related to the role of each object. Indeed, and as one may observe in the
effective action to be built below, any field whose last subscript is 1 behaves
as
a ghost associated to some of the gauge symmetries of the
original action, if the last identifier is a 2 the object is an antighost while
if it is 3 the field is a Lagrange multiplier accompanying a gauge fixing
condition.

Let us now turn our attention to the construction of the effective action.
Since the theory under study has a vanishing Hamiltonian its effective action
is given by
[10]:
$$
\align
S_{eff}=\int^{tf}_{ti}dt[&\pi^{i}\dot{A}_i+\mu^1 \dot{C}_1
+\mu^{1ij}_{(0)}\dot{C}_{1ij}^{(0)}+\mu^{11j}_{(1)}\dot{C}_{11j}^{(1)}
+\mu^{111}_{(2)}\dot{C}_{111}^{(2)}\\
&\wh{\delta}(\lambda_1\mu^{1}+\lambda_{1ij}^{0(0)}\mu^{1ij}_{(0)}+
\lambda_{11i}^{1(1)}\mu^{11i}_{(1)}+\lambda_{111}^{2(2)}\mu^{111}_{(2)})
+L_{GF+FP}], \tag 14
\endalign
$$
where
$$
\align
L_{GF+FP}=&\wh{\delta}(C_2\chi_2+C^{(0)}_{2}\chi_2^{(0)})+
\wh{\delta}(\sum^3_{M=1}C^{(1)}_{M2}\chi^{(1)}_{M2}+
\sum^3_{M,N=1}C^{(2)}_{MN2}\chi^{(2)}_{MN2})+\\
&\wh{\delta}(\lambda_{12}^{0(1)}\Lambda_2^{0(1)}+
\sum^3_{M=1}\lambda_{1M2}^{0(2)}\Lambda_{M2}^{0(2)}+
\lambda_{112}^{1(2)}\Lambda_{2}^{1(2)})+ \\
&\wh\delta(\theta_{12}^{0(1)}\Theta_{2}^{0(1)}+
\sum^3_{M=1}\theta_{1M2}^{0(2)}\Theta_{M2}^{0(2)}+
\theta_{112}^{1(2)}\Theta_{2}^{1(2)})
,\tag 15
\endalign
$$
is the sum of the generalizations of the Fadeev-Popov and gauge fixing
terms. In (15) $\chi_2$, $\chi^{(0)}_2$ are the primary gauge fixing
functions associated  to the constraints (4a) and (4b), while
$\chi^{(1)}_{M2}$,
$\chi^{(2)}_{MN2}$, $\Lambda_2^{0(1)}$, $\Lambda_{M2}^{0(2)}$,
$\Lambda_{2}^{1(2)}$  $\Theta_2^{0(1)}$ ,$\Theta_{M2}^{0(2)}$ and
$\Theta_{2}^{1(2)}$ are gauge fixing functions which must fix the longitudinal
part of the fields in the non minimal sector. As usual, the BRST transformation
for the canonical variables ($Z$) is given by
$$
\wh{\delta}Z=(-1)^{\epsilon_z}\{ Z,\Omega \},\tag 16
$$
where $\epsilon_z$ is the grassmanian parity of $Z$, while the BRST
transformation of the variables belonging to the non minimal sector are fixed
by imposing the closure of the charge as discussed in [10]. Their explicit
form for this system may be read from the expressions we give below for the
general D-dimensional case.

Let us now turn to the general case. For D$\geq$ 6
once again the action (1)  can be written as a constrained canonical action
with vanishing Hamiltonian:

$$ S=\int d^Dx (\dot{A}_i^a \pi^{ia} + A_0^a\phi^a
+B_{0 i_1 \cdots i_{D-1}}^a\Phi^{i_1\cdots i_{D-1} a}).
\tag 17
$$

This time the constraints are defined as follows
$$
\align
&\phi^a \equiv (\nabla_i\pi^i)^a=\partial_i\pi^{ia}+(A_i\times
\pi^i)^a=\partial_i\pi^{ia}+f^{abc}A_i^b\pi^{ic}=0,\tag 18a\\
&\Phi^{i_1\cdots i_{D-1} a} \equiv
\epsilon^{i_1\cdots i_{D-1}}(F_{i_{D-2} i_{D-1}}^a)=0.\tag 18b
\endalign
$$

Clearly, the constraints have the same structure seen in the 4 [9] and 5
dimensional cases; in fact, the $ \phi^a $ constraints are exactly the same.
The Poisson Bracket algebra of these constraints is still first class and
explicitly given by
$$
\align
&\{\phi^a(x),\phi^b(x')\} =f^{abc}\phi^c(x)\delta^{D-1}(x-x'),\tag19a \\
&\{\Phi^{i_1\cdots i_{D-1} a}(x),\Phi^{j_1\cdots j_{D-1} b}(x')\} =0,\tag19b\\
&\{\phi^a(x),\Phi^{i_1\cdots i_{D-1} a}(x')\}
=f^{abc}\Phi^{i_1\cdots i_{D-1} a}(x)\delta^{D-1}(x-x') .\tag19c
\endalign
$$

As mentioned the D-dimensional BF theories  have D-3 stages of reducibility
which are defined through the following operators
$$
a^{(1)}\equiv a^{(1) j_1 \cdots j_{D-4}a}_{i_1 \cdots i_{D-3}}=
\delta^{j_1}_{[i_1} \delta^{j_2}_{i_2} \cdots \delta^{j_{D-4}}_{i_{D-4}}
\nabla^a_{i_{D-3}]}, \tag 20a
$$
$$
a^{(2)}\equiv a^{(2) j_1 \cdots j_{D-5}a}_{i_1 \cdots i_{D-4}}=
\delta^{j_1}_{[i_1} \delta^{j_2}_{i_2} \cdots \delta^{j_{D-5}}_{i_{D-5}}
\nabla^a_{i_{D-4}]}, \tag 20b
$$
\centerline{$ \cdots$}
$$
a^{(D-4)}\equiv a^{(D-4) k a}_{ij}=
\delta^{k}_{[i} \nabla^a_{j]}, \tag 20c
$$
$$
a^{(D-3)}\equiv a^{(D-3) a}_m = \nabla^a_m. \tag 20d
$$

For these theories [10] the minimal sector of the extended phase space is
itself
divided in two parts: one composed by
the fields associated to the  irreducible constraints which are the same as
in the lower dimensional case, see (9b), and one composed by the fields
associated to the reducible constraints which are the following canonical
pairs that generalize (9c) (notice the geometric structure given by the
 space-like indices):
$$ \align
&(C_{1 i_{1}\dots i_{D-3}}^{(0)},\mu^{1 i_{1}\dots i_{D-3}}_{(0)}) \\
&(C_{11 i_{1}\dots i_{D-4}}^{(1)},\mu^{11 i_{1}\dots i_{D-4}}_{(1)})
\tag21a
\endalign
$$
\centerline{$ \cdots$}
$$
(C_{\underbrace{1 \dots 1}_p i_{1}\dots i_{D-2-p}}^{(p-1)},
\mu^{\overbrace{1 \dots 1}^p i_{1}\dots i_{D-2-p}}_{(p-1)})
\tag21b
$$
\centerline{$ \cdots$}
$$
(C_{\underbrace{1 \dots 1}_{D-2}}^{(D-3)},
\mu^{\overbrace{1 \dots 1}^{D-2}}_{(D-3)}).
\tag21c
$$

To simplify the reading we introduce the following
compact notation to be used hereon: a bracketed subscript (or superscript)
will refer  to the number  of "ones" which label an object while the
$(D-2-p)$ space indices will be represented by a greek multi-index. For
example:
$$
\align
&C_{[p] \epsilon} \equiv C_{\underbrace{1 \dots 1}_p i_{1}\dots
i_{D-2-p}}^{(p-1)} \\
&\mu^{[p] \epsilon} \equiv \mu^{\overbrace{1 \dots 1}^p
i_{1}\dotsi_{D-2-p}}_{(p-1)},
\   p=1,2,\dots,D-2 . \tag22
\endalign
$$
Additionaly we also use the following concise notation for
the reducibility operators,
$$
a^{(p)}\equiv a^{(p) \epsilon}_{\nu} \equiv
{\underbrace{\delta^{j_1}_{[i_1}
\delta^{j_2}_{i_2} \cdots \delta^{j_{D-2-p}}_{i_{D-2-p}}}_{(p-1) delta
factors}}
\nabla^a_{i_{D-1-p}]}, \  p=1,2,\dots,D-3 . \tag23
$$

With this notation the general expression for the off shell nilpotent
BRST operator is given by
$$
\align
\Omega^{\D}=<&
C_1\phi+C_{[1] \mu}\Phi^{\mu}
+\sum^{\D -2}_{p=2}C_{[p] \mu}[\delta\nabla]^{\mu}_{\nu} \mu^{[p-1] \nu})\\
&-{1\over{2}}C_1(C_1\times \mu^1)-\sum^{\D
-2}_{p=1}C_{[p]\mu}(C_1\times\mu^{[p]\mu})\\
&+\sum^{p+q+1=\D-2}_{p+q+1=3}(2-\delta_{pq})A_{pq}C_{[p+q+1]\mu}(\mu^{[p]\mu\nu}
\times\mu^{[q]\alpha})\epsilon_{\nu\alpha}> .
\tag 24
\endalign
$$
This expression for the BRST charge has
essentially the same structure that we observed in the five dimensional case.
In fact we recognize the Yang Mills like terms, the
$C_{[p]}[\delta\nabla]\mu^{[p-1]} $ terms and the new polinomial
$C_{[p+q+1]}(\mu^{[p]}\times\mu^{[q]})$ terms that ensure the off-shell
nilpotency of the charge. The unknown coefficients can be explicitly calculated
through a recurrence relation which for $D\ge5$ and $p \ge q$ yields:

$$
(D-2-p-q)!  A_{p,q-1}+(-1)^q(D-2-p)! A_{p+1,q-1}=0 , \tag25a
$$
$$
(-1)^{p+1}(D-2-p-q)! A_{p,q-1}+(-1)^{D-1}(q+1)(D-2-q)! A_{p,q} =0 , \tag25b
$$

$$
\align
A_{D-4,1}&={(-1)^D\over{2(D-3)!}} , \tag25c\\
A_{1,1}&=-{(D-4)!\over{4}} .  \tag25d
\endalign
$$

It only remains to construct the non minimal sector of
the phase space, the BRST transformation properties of their fields and the
effective action. The fields in the minimal sector associated to the
irreducible constraints  are those given by Table 1. To display the complete
set of fields in the non-minimal sector  associated to the  reducible
constraints one has to generalize the tree diagrams of figures 1 and 2.
The non minimal $C$ fields are given by:
$$
C^{(p)}_{\underbrace{SJ \cdots I}_{(p+1)-subscripts}}  \   S,J,...,I=1,2,3; \
\  p=0,1,\dots, D-3 \tag26
$$
where at least one of the capital subscripts must take the values 2 or 3.

And the $\lambda$ and $\theta$ fields which complete the set of local
coordinates of the extended phase space:
$$ \align
&\lambda^{r(s)}_{[r+1] \underbrace{M\cdots Q}_{s-r} \epsilon} \equiv
\lambda^{r(s)}_{\underbrace{1\cdots 1}_{r+1}
\underbrace{M\cdots Q}_{s-r} \epsilon} \tag27a \\
&\theta^{r(s)}_{[r+1] \underbrace{M\cdots Q}_{s-r} \epsilon}
\equiv\theta^{r(s)}_{\underbrace{1\cdots 1}_{r+1} \underbrace{M\cdots Q}_{s-r}
\epsilon} \tag27b \endalign
$$
$r=0,1,\dots,D-3; s=0,\dots,D-3-r; \ M,Q,P =1,2,3$

Next we define the BRST transformation rules for all the fields. In the
minimal sector, the BRST transformation is simply given by:
$$
\wh{\delta}Z = (-1)^{\epsilon_{z}}\{Z,\Omega\} . \tag28
$$
In the non-minimal sector the BRST transformation laws must be calculated to
ensure their off-shell nilpotency (i.e.
$\wh{\delta}\wh{\delta}$(anything)=0). To achieve this goal  we
need to introduce the
transverse-longitudinal (T+L) decomposition of geometrical objects with
respect to the reducibility operators ($a^{(p) \epsilon}_{\mu}$).

Given a lower multi-indexed geometrical object $V_{\epsilon}$ its T+L
decomposition is given as:
$$
\align
&V_{\epsilon}=V^T_{\epsilon}+a^{(p) \mu}_{\epsilon}V_{\mu} \tag29a \\
&A^{(p) \epsilon}_{\mu}V^T_{\epsilon}=0 \tag29b \\
&V^L_{\mu}=A^{(p) \epsilon}_{\mu}V_{\epsilon} . \tag29c
\endalign
$$
Similarly, for upper multi-indexed objects ($V^{\rho}$), the T+L
decomposition is defined through
$$
\align
&V^{\rho}=V^{T \rho}+A^{(p) \rho}_{\mu}V^{L \mu} \tag30a \\
&a^{(p) \mu}_{\rho}V^{T \rho}=0 \tag30b \\
&V^{L \rho}=a^{(p) \rho}_{\mu}V^{\mu} \   p=1,\dots,D-3 . \tag30c
\endalign
$$
It is important to remark that there is always possible to find a set
of operators $A^{(p) \mu}_{\nu} (p=1, \dots ,D-3)$ such that the above
decompositions  are unique.
Next, let  $t$ stand for any variable of the non-minimal sector. The BRST
transformation rules are given by
$$
\wh{\delta} t^{(i+j)}_{\underbrace{M\cdots Q}_{i-1}
2[j] \epsilon}=
t^{(i+j)}_{\underbrace{M\cdots Q}_{i-1}
3[j]\epsilon}+a^{(i+j+1) \mu}_{\epsilon}t^{(i+j+1)}_{\underbrace{M\cdots
Q}_{i-1} 2[j+1] \mu} \tag31a
$$

$$
\wh{\delta}t^{(i+j)}_{\underbrace{M\cdots Q}_{i-1}
3[j] \epsilon}=
-\wh{\delta}a^{(i+j+1) \mu}_{\epsilon}t^{(i+j+1)}_{\underbrace{M\cdots Q}_{i-1}
2[j+1] \mu}-a^{(i+j+1)
\mu}_{\epsilon}\wh{\delta}t^{(i+j+1)}_{\underbrace{M\cdots Q}_{i-1}
2[j+1] \mu} \tag31b
$$
for $1 \leq i \leq D-3; 1 \leq i+j \leq D-4 $ (If the object is upper indexed
one must change the $a^{(p)\mu}_{\nu}$ operator for the $A^{(p)\mu}_{\nu}$
one).

For the last level of reducibility one finds
$$
\align
&\wh{\delta} t^{(D-3)}_{\underbrace{M\cdots Q}_{D-3}
2 \epsilon}= t^{(D-3)}_{\underbrace{M\cdots Q}_{D-3}3 \epsilon} \tag31c \\
&\wh{\delta} t^{(D-3)}_{\underbrace{M\cdots Q}_{D-3}3 \epsilon}= 0 \tag31d \\
&\wh{\delta} t^{(D-3)}_{\underbrace{M\cdots Q}_{i}
2[j] \epsilon}=t^{(D-3)}_{\underbrace{M\cdots Q}_{i}
2[j] \epsilon} \tag31e \\
&\wh{\delta} t^{(D-3)}_{\underbrace{M\cdots Q}_{i}
3[j] \epsilon}= 0 ; \  i+j=D-3 . \tag31f
\endalign
$$
For the $\lambda$ and $\theta$ fields which define
the begining of a tree diagram the transformation laws are
$$
\align
&\wh{\delta}
\lambda^{i(i)}_{[i+1]\mu}=
\theta^{i(i)}_{[i+1]\mu}
+a^{(i+1)\rho}_{\mu}
\lambda^{i(i+1)}_{[i+2]\rho} \tag32a \\
&\wh{\delta}
\theta^{i(i)}_{[i+1]\mu}=
-\wh{\delta}a^{(i+1)\rho}_{\mu}
\lambda^{i(i+1)}_{[i+2]\rho}
-a^{(i+1)\rho}_{\mu}
\wh{\delta}\lambda^{i(i+1)}_{[i+2]\rho} \tag32b \\
&i=0,\dots D-4
\endalign
$$
and finally:
$$
\align
&\wh{\delta}
\lambda^{D-3(D-3)}_{[D-2]}=\theta^{D-3(D-3)}_{[D-2]} \tag32c
\\ &\wh{\delta}\theta^{D-3(D-3)}_{[D-2]}=0 . \tag32d
\endalign $$

With the  BRST transformation rules given above, the effective action is built
as a generalization of (14) as follows [10]:
$$
Seff=\int^{tf}_{ti}dt[\pi^{i}\dot{A}_i+\mu^1
\dot{C}_1+\sum^{D-3}_{p=0}\mu^{[p+1]} \dot{C}_{[p+1]}
+\wh{\delta}(\sum^{D-3}_{p=0} \lambda^{p(p)}_{[p+1]}\mu^{[p+1]})
+L_{GF+FP}] . \tag33
$$
this time the Fadeev-Popov + gauge fixing Lagrangian is given by a longer
expression which includes enough terms as to completely fix the gauge
associated to the reducible constraints.
$$
\align
L_{GF+FP}=&\wh{\delta}(C_2\chi_2+C^{(0)}_{2}\chi_2^{(0)}+\\
&\wh{\delta}(\sum^3_{M=1}C^{(1)}_{M2}\chi^{(1)}_{M2}+
\dots+\sum^{3}_{M,N,\dots P=1}
C^{(D-3)}_{\underbrace{MN\dots P}_{D-2}2}\chi^{(D-3)}_{MN\dots P2}) \\
&\wh{\delta}(\sum^{D-4}_{r=0}
\sum^{D-3}_{s=r+1}\sum^3_{M,N,\dots,P=1}
\lambda_{[r+1]\underbrace{MN \dots P}_{s-r-1}2}^{r(s)}
\Lambda_{\underbrace{MN \dots P}_{s-r-1}2}^{r(s)})+ \\
&\wh{\delta}(\sum^{D-4}_{r=0}
\sum^{D-3}_{s=r+1}\sum^3_{M,N,\dots,P=1}
\theta_{[r+1]\underbrace{MN \dots P}_{s-r-1}2}^{r(s)}
\Theta_{\underbrace{MN \dots P}_{s-r-1}2}^{r(s)}) \\.
\tag34
\endalign
$$

The admissible set of gauge choices includes only the ones which fix the
longitudinal part of the associated fields. Within this set the modified BFV
approach [10-11] ensures that the functional integral of the theory (defined
as the sum over histories of
$exp(-S_{eff})$ with unit weight) is locally independent of the gauge choice
provided the following boundary condition holds:
$$
[\Omega-\pi^{\rho}{{\partial\Omega}\over{\partial\pi^{\rho}}}-\sum^{D-3}_{p=1}\mu^{[1]
\rho}
{{\partial\Omega}\over{\partial\mu^{[1]\rho}}}]\mid^{t_{fin}}_{t_{in}}=0 .
\tag35
$$

{}From the obtained  expression for the BRST charge $\Omega$ one verifies that
the partition function {\bf $Z$} is {\bf independent of the gauge coupling
constant} ($e^2$). In fact the
effective action may be written as a linear homogeneous quantity in the
following set of variables:
$$
\pi^i, \mu, C_{[1]},C_{[2]},\dots,C_{[D-2]} \tag36a
$$
and
$$
\lambda^{r(p)}_{[p+1]},\theta^{r(p)}_{[p+1]}, \  p=0,\dots, D-3
\  ,r=0,\dots, D-3 \tag36b
$$
plus the gauge fixing terms. The gauge coupling constant may then be
absorbed by redefining the set (36) together with a change in the gauge
fixing choice provided $e^2\ne 0$. Since the partition function {\bf $Z$} is
independent of the gauge fixing condition, {\bf $Z$} does not depend
on the coupling constant. This
property allows {\bf $Z$} to be evaluated by going to the limit of very small
$e^2$ where the path integral is dominated by the classical minima. We may
then conclude that the semi-classical limit is exact.

The independence on the
coupling constant  is  a common link between many topological gauge
theories (see [13],[16],[18]).
For Witten's topological theory it was proved in [13]  using the fact that
the Lagrangean for such theory is of the form $\{\Omega,\Psi\}$. In
the case of the non abelian BF theories  the effective
Lagrangean has not this structure [9] but our argument based
on the form of the $\Omega$ operator allows to reach the same conclusion.

\vskip 3cm
\noi
{\bf REFERENCES}
\vskip .3cm

\item{[1]}G. T. Horowitz, {\it Commun. Math.Phys.}{\bf125} (1989) 417.
\item{[2]}G. T. Horowitz and M. Srednicki, {\it Commun. Math.Phys.}{\bf130}
(1990) 83.
\item{[3]}M. Blau and G. Thompson, {\it Ann.Phys.} {\bf 205}
(1991) 130; D. Birmingham, M. Blau, M. Rakowski and G. Thompson, {\it Phys.
Rep.}
{\bf 209} (1991) 129.
\item{[4]}B. Broda, {\it Phys. Lett.}
{\bf B254} (1991) 111.B. Broda, {\it Phys. Lett.}
{\bf B280} (1992) 47.
\item{[5]}M. Blau and G. Thompson, {\it Phys. Lett.} {\bf 255}
(1991) 535.
\item{[6]}M. Abud and G. Fiore,{\it Triestre Preprint SISSA 159/91 EP and
Napoli Preprint INFN, NA IV 21/91}.
\item{[7]} J.C. Wallet, {\it Phys. Lett.} {\bf 235} (1990) 71.
\item{[8]}C. Lucchesi, O. Piguet and S.P. Sorella, {\it Nucl. Phys.} {\bf B395}
(1993) 325.

\item{[9]}M. I. Caicedo and A. Restuccia, {\it Phys. Lett.} {\bf B307 } (1993)
77.
\item{[10]}M. I. Caicedo and A. Restuccia, {\it Class. Quan. Grav} {\bf 10 }
(1993) 833.
\item{[11]}M. I. Caicedo, {\it Doctoral Thesis, Universidad Sim\'on
Bol\'ivar}  (1993); Preprint {\bf SB/F/93-221}.
\item{[12]}E. S. Fradkin and G. A. Vilkovisky, {\it Phys. Lett.} {\bf B55}
(1975)
224; CERN report TH-2332, (1977); I. Batalin and E. Fradkin, {\it Phys. Lett.}
{\bf B122} (1983) 157; {\it Phys. Lett.} {\bf B128} (1983) 307; {\it Ann Inst
Henri Poincar\'e} {\bf 49} (1988) 215.
\item{[13]}E. Witten,{\it Commun. Math. Phys.} {\bf 117} (1988) 353.
\item{[14]}J. M. F. Labastida and M. Pernici, {\it Phys. Lett.} {\bf B212}
(1988) 56; F.De Jonghe and S.Vandoren, {\it Phys. Lett.} {\bf B324} 328.
\item{[15]}L. Baulieu and I. M. Singer, {\it Nucl. Phys.} (Proc.
Suppl.) {\bf 5B} (1988) 12; Y. Igarashi, H. Imai, S. Kitakado and H. So, {\it
Phys. Lett.} {\bf B227} (1989) 239; C. Arag\~{a}o and L. Baulieu, {\it Phys.
Lett.} {\bf B275} (1992) 315.
\item{[16]}R.Gianvittorio, A.Restuccia and J.Stephany, Preprint {\bf
SB/F/94-225} {\it Phys Lett.B} to appear (hep-th/9410123).
\item{[17]}N. Seiberg and E. Witten, {\it hep-th/9407087, hep-th/9408099}.
\item{[18]}L.F.Cugliandolo,G.Lozano and F.A.Schaposnik, {\it Phys.Lett.} {\bf
B234 } (1990) 52.

\vfil\eject
\end